\documentclass[11pt]{article}
\pdfoutput=1 
\usepackage{amssymb}
\usepackage{amsmath}
\usepackage{amstext}
\usepackage{mathrsfs}
\usepackage{multirow}
\usepackage{graphicx}
\usepackage{epsfig}
\usepackage{verbatim} 
\usepackage{fancybox}
\usepackage{color}
\usepackage{mathdots}
\usepackage{anyfontsize}
\usepackage{ulem}
\usepackage{enumitem}
\usepackage{caption}
\usepackage{amsthm}
\usepackage{subfig}
\usepackage{mathtools}
\usepackage{youngtab}
\usepackage{braket}
\usepackage{bbm}
\usepackage{parskip}
\usepackage[numbers,sort&compress]{natbib}
\usepackage{ytableau}
\usepackage[utf8]{inputenc}
\usepackage{tikz}
\usetikzlibrary{arrows.meta}
\usepackage{simplewick}
\usepackage{overpic}
\usetikzlibrary{positioning, trees,decorations, decorations.markings, decorations.pathmorphing, arrows, shapes.geometric}

\newcommand{\Comment}[1]{{}}
\definecolor{darkblue}{rgb}{0.15,0.35,0.55}
\definecolor{reddish}{rgb}{0.65, 0.2, 0.2}
\usepackage[linktocpage=true]{hyperref}
\hypersetup{
colorlinks=true,
citecolor=darkblue,
linkcolor=reddish,
urlcolor=darkblue,
pdfauthor={},
pdftitle={},
pdfsubject={}
}

\usepackage{cleveref}

\newcommand{\be}{\begin{equation}}
\newcommand{\ee}{\end{equation}}
\newcommand{\nn}{\nonumber}

\renewcommand{\vec}[1]{\boldsymbol{#1}}

\numberwithin{equation}{section}

\begin{document}

\renewcommand{\thefootnote}{\fnsymbol{footnote}}
~
\vspace{1.75truecm}
\thispagestyle{empty}
\begin{center}
{\LARGE \bf{
Moduli Bounds from Spin-2 Sum Rules
}}
\end{center} 

\vspace{1cm}

\begin{center}
{\fontsize{13.5}{18}\selectfont
Francesco Bertucci${}^{\rm a,}$$^{}$\footnote{E-mail: \Comment{\href{mailto:fbertucc@olemiss.edu}}{\tt fbertucc@olemiss.edu}}, James Bonifacio${}^{\rm a,}$$^{}$\footnote{E-mail: \Comment{\href{mailto:bonifacio@phy.olemiss.edu}}{\tt bonifacio@phy.olemiss.edu}}, and Kurt Hinterbichler${}^{\rm b,}$$^{}$\footnote{E-mail: \Comment{\href{mailto:kurt.hinterbichler@case.edu}}{\tt kurt.hinterbichler@case.edu}} 
}
\end{center}
\vspace{.4truecm}

 \centerline{{\it ${}^{\rm a}$Department of Physics and Astronomy,}}
 \centerline{{\it University of Mississippi, University, MS 38677}} 
 
   \vspace{.3cm}

 \centerline{{\it ${}^{\rm b}$CERCA, Department of Physics,}}
 \centerline{{\it Case Western Reserve University, 10900 Euclid Ave, Cleveland, OH 44106}} 
 
  \vspace{.3cm}

\vspace{1cm}
\begin{abstract}
\noindent
Mirbabayi and Villadoro have provided strong numerical evidence for a universal upper bound on the mass of the lightest scalar field in gravitational Kaluza--Klein (KK) theories. This bound states that the lightest KK graviton, with mass $m_{1}$, must couple to at least one scalar with mass $m_{\rm sc}$ such that $(m_{\rm sc} /m_{1})^2 \leq 4/3$. We give a proof of this bound using the approach of the conformal bootstrap applied to sum rules for massive spin-2 scattering amplitudes. We additionally show that each heavier KK graviton, with mass $m_n$, must couple to at least one scalar with mass $m_{\rm sc}$ such that $(m_{\rm sc} /m_{n})^2 < 36/25$.
\end{abstract}

\newpage
\setcounter{tocdepth}{2}
\tableofcontents
\renewcommand*{\thefootnote}{\arabic{footnote}}
\setcounter{footnote}{0}

\section{Introduction}

Our universe may be higher dimensional, yet still appear to us as four dimensional, if the extra dimensions are curled up into a small internal manifold.  This Kaluza--Klein (KK) idea is an essential ingredient in the majority of phenomenological models relying on a higher-dimensional setup.  A universal feature of KK models is the existence of towers of massive particles of spin $\leq 2$, whose masses are given by the eigenvalues of various Laplacians acting on the internal manifold (see \cite{Hinterbichler:2013kwa} for a summary).  In particular, there will be a tower of massive spin-2 particles, the KK gravitons, on top of the massless graviton. The mass of the lightest of these, $m_1$, sets the KK scale, which is the typical energy scale at which extra-dimensional physics becomes relevant.  

In addition to the KK gravitons coming in at the scale $m_1$, another universal feature of KK theories is the presence of moduli.  These are scalar fields, coupled with gravitational strength, that correspond to deformations in the overall size or structure of the compact space, and they are often massless or much lighter than the KK scale.  We do not observe light, gravitationally coupled scalar fields, and a natural explanation for this is that the moduli have been made massive by some stabilization mechanism that stiffens the problematic deformations of the internal manifold \cite{Dine:1985rz,Goldberger:1999uk,Giddings:2001yu,Kachru:2003aw,Balasubramanian:2005zx,McAllister:2023vgy}.\footnote{Another possibility is a screening mechanism that hides the scalar in the presence of matter \cite{Vainshtein:1972sx,Khoury:2003aq,Hinterbichler:2010es,Babichev:2013usa}, but this is harder to realize in a top-down KK setup.}

In \cite{Mirbabayi:2026saz}, Mirbabayi and Villadoro propose an interesting bound, putting a limit on how massive a stabilization mechanism can make the moduli.  The bound of Mirbabayi and Villadoro, which they support with strong numerical evidence, states that there must exist a scalar of mass $m_{\rm sc}$ that couples to the lightest KK graviton such that
\be \left( m_{\rm sc} \over m_{1}\right)^2 \leq {4\over 3}\,.\label{mainbounde}\ee

The assumptions going into this bound are minimal: it is assumed that the spectrum is discrete and contains only spins up to 2, that the tree-level 4-point amplitude of the lightest massive spin-2 particle is non-zero, and that the high-energy behavior of this 4-point amplitude behaves gravitationally as it would in a KK theory, meaning that it does not grow faster with energy than $E^2$; this is the behavior of the tree-level amplitude in Einstein gravity, to which a KK theory  must revert at high energies where it becomes higher dimensional.  Because of the paucity of these assumptions, the bound \eqref{mainbounde} is insensitive to the details of the internal manifold or the stabilization mechanism --  it does not matter if the internal manifold is smooth, what structures it has, whether it forms a warped or direct product with the four-dimensional space, or anything else.  All that matters is that there is a four-dimensional flat space  description in terms of a discrete set of fields with spins $\leq 2$.  

In fact, it need not be a KK theory at all.  The bound \eqref{mainbounde} applies to any theory containing a massive spin-2 particle with parity-even two-derivative interactions and no couplings to higher spins, so long as the assumptions about the 4-point amplitude and its $E^2$ growth are accepted.  Note that the $E^2$ growth assumption is essential: effective field theories of massive spin-2 fields, including couplings to gravity, can exist without any scalars \cite{deRham:2010kj,Hassan:2011zd,Hinterbichler:2012cn}, but they always have tree-level amplitudes with stronger high-energy growth of at least $E^6$ \cite{Schwartz:2003vj, Bonifacio:2018vzv,Bonifacio:2018aon}.

We will show that a proof of the bound \eqref{mainbounde} follows from sum rules on massive spin-2 scattering derived in \cite{Bonifacio:2019ioc}, following studies of massive spin-2 amplitudes and their high-energy behavior in \cite{Bonifacio:2018vzv,Bonifacio:2018aon,Bonifacio:2019mgk}.  
When applied to compactifications on a closed manifold, these sum rules can be used to derive constraints on the spectral data of manifolds and arise from expanding quadruple overlap integrals of eigenfunctions in multiple ways \cite{Bonifacio:2019ioc,Bonifacio:2020xoc}, analogous to the manner in which the conformal bootstrap derives constraints from equating different OPE channel expansions of a four-point correlator. See \cite{Bonifacio:2021msa,Bonifacio:2021aqf,Kravchuk:2021akc,Gesteau:2023brw,Bonifacio:2023ban, Radcliffe:2024jcg, Adve:2025rvf, Adve:2025sld, Bertucci:2026fjh} for further results from this approach.  However, the sum rules we use, which were called ``bottom-up'' sum rules in \cite{Bonifacio:2019ioc}, are independent of geometry or KK compactification, and apply to any amplitudes under the assumptions stated above. The sum rules can also be obtained from the constraints given in \cite{Mirbabayi:2026saz}, which are derived in a similar way.

In addition to \eqref{mainbounde}, there are other similar bounds that can be derived from the sum rules.  We show that for any spin-2 particle with mass $m_n$, not necessarily the lightest one, there must exist a scalar with mass $m_{\rm sc}$ that couples to it such that
\be \left(m_{\rm sc} \over m_{n}\right)^2 < {36\over 25}\,.\label{newboundee}\ee
For the lightest massive spin-2 particle, this is slightly weaker than \eqref{mainbounde}, but it also applies to all the other massive spin-2 particles in the spectrum.

\section{Review of massive spin-2 sum rules}

We first briefly review the derivation of the bottom-up sum rules from \cite{Bonifacio:2019ioc}.  Consider a four-dimensional theory containing a massive spin-2 field $h_{\mu \nu}^{\star}$ with mass $m_{\star}$.  The central assumption is that the tree-level four-point scattering amplitude $h^\star h^\star \rightarrow h^\star h^\star$ is non-zero and grows at most like $ E^2$ for $m_\star \ll E$, 
where $E$ is the center-of-mass energy.\footnote{In a full quantum theory, there is an additional energy scale $\Lambda$ at which loops or higher-spin particles become important and we must assume that $E \ll \Lambda$.}
Other assumptions are that the cubic interactions $h^\star h^\star X$, with $X$ any particle, are all parity-even (i.e., the interactions do not make use of the Levi--Civita symbol)
and contain at most two derivatives.
Finally, it is also assumed that all the $X$ have spin $\leq 2$ and thus, other than $h_{\mu \nu}^{\star}$ itself, the particle $X$ can be:
\begin{itemize}
\item  A massless graviton $h_{\mu \nu}^0$.
\item One of a collection of other massive spin-2 fields $h_{\mu \nu}^a$, labeled by an index $a$, and with masses $m_a$.
\item One of a collection of spin-1 fields $A_{\mu}^i$, labeled by an index $i$, and with masses $m_i$.
\item One of a collection of scalar fields $\phi^{\mathcal{I}}$, labeled by an index $\mathcal{I}$, and with masses $m_{\mathcal{I}}$.
\end{itemize}
  We can ignore fermionic fields since they cannot appear in cubic couplings with two bosons, so there is no restriction on the fermionic content or their spins.
  
  The most general on-shell $h^\star h^\star X$ vertices satisfying these assumptions can then be written as
\begin{align}
\mathcal{V}(1_{h^{\star}}, \, 2_{h^{\star}}, \, 3_{h^{\star}}) &=i a_1 m_{\star}^2 \epsilon_{12}   \epsilon_{13}   \epsilon_{23}  +i a_2 \left[ \epsilon_{23}^2 (\epsilon_1 \cdot p_2)^2+ \epsilon_{13} ^2 (\epsilon_2 \cdot p_3)^2+ \epsilon_{12}^2 (\epsilon_3 \cdot p_1)^2\right]  \nn \\ 
& \!\!+i a_3 \left[  \epsilon_{13}  \epsilon_{23}   (\epsilon_1 \cdot p_2)  (\epsilon_2 \cdot p_3)+ \epsilon_{12}   \epsilon_{23}  (\epsilon_1 \cdot p_2) (\epsilon_3 \cdot p_1)+ \epsilon_{12} \epsilon_{13}  (\epsilon_2 \cdot p_3 ) (\epsilon_3 \cdot p_1) \right] ,  \\
\mathcal{V}(1_{h^{\star}}, \, 2_{h^{\star}}, \, 3_{h^0}) & = i b_1 \epsilon_{12}^2 (\epsilon_3 \cdot p_1)^2 +i b_2 \epsilon_{12} ( \epsilon_3 \cdot p_1) \big[\epsilon_{23}  (\epsilon_1 \cdot p_2)+\epsilon_{13}  (\epsilon_2 \cdot p_3) \big] \nn \\
&+i b_3 \big[ \epsilon_{23} (\epsilon_1 \cdot p_2)+\epsilon_{13} (\epsilon_2 \cdot p_3) \big]^2  , \\
\mathcal{V}(1_{h^{\star}}, \, 2_{h^{\star}}, \, 3_{{\phi}^{\mathcal{I}}}) &  = i m^2_{\star} c_{1,\mathcal{I}} \epsilon_{12}^2 +ic_{2,\mathcal{I}} \, \epsilon_{12} (\epsilon_1 \cdot p_2)( \epsilon_2 \cdot p_3),  \\
\mathcal{V}(1_{h^{\star}}, \, 2_{h^{\star}}, \, 3_{A^i}) & =m_{\star} d_{i}  \epsilon_{12}  \big[ \epsilon_{23} (\epsilon_1 \cdot p_2)-\epsilon_{13} (\epsilon_2 \cdot p_3) \big],\\ 
\mathcal{V}(1_{h^{\star}}, \, 2_{h^{\star}}, \, 3_{h^a}) &=i m_{\star}^2 e_{1,a}  \epsilon_{12}  \epsilon_{13}  \epsilon_{23}  +i e_{2,a} \epsilon_{12}^2   (\epsilon_3 \cdot p_1)^2 +i e_{3, a}\left[ \epsilon_{12} \epsilon_{23}  (\epsilon_1 \cdot p_2) (\epsilon_3 \cdot p_1)+ \epsilon_{12}  \epsilon_{13}  (\epsilon_2 \cdot p_3)  (\epsilon_3 \cdot p_1) \right]  \nn \\ & +i e_{4,a} \left[ \epsilon_{23}^2 (\epsilon_1 \cdot p_2)^2+ \epsilon_{13}^2 (\epsilon_2 \cdot p_3)^2\right] + i e_{5,a} \epsilon_{13}  \epsilon_{23}  (\epsilon_1 \cdot p_2)(\epsilon_2 \cdot p_3), 
\end{align}
where we write the symmetric transverse traceless tensor polarizations as $\epsilon_{a \mu} \epsilon_{a \nu}$ with $\epsilon_a \cdot \epsilon_a =0$ and use the shorthand $\epsilon_{ab} = \epsilon_{a } \cdot \epsilon_b$. The coupling constants $a_1$, $a_2$, $a_3$, $b_1$, $b_2$, $b_3$, $c_{1, \mathcal{I}}$, $c_{2, \mathcal{I}}$,  $d_i$,  $e_{1, a}$, $e_{2, a}$, $e_{3, a}$, $e_{4, a}$, and $e_{5 , a}$ are all real. Gauge invariance and the S-matrix equivalence principle furthermore require that in the presence of gravity we have $b_1 = b_2/2=2/M_P$, where $M_P$ is the Planck mass in four dimensions.

The aim is to find sum rules that must be satisfied by the cubic coupling constants if the $h^\star h^\star \rightarrow h^\star h^\star$ tree amplitude is not to grow faster than $E^2$.  This amplitude contains the exchange contributions from exchanging all possible $X$ particles, which can be constructed from the above general cubic vertices, as well as a contact contribution from a possible $h^\star h^\star h^\star h^\star$ vertex.
The most general on-shell $h^\star h^\star h^\star h^\star$ vertex  with up to six derivatives is given by a 95-parameter polynomial in the contractions $\epsilon_{ab}$, $\epsilon_a \cdot p_b$, and $p_a\cdot p_b$.\footnote{The 95 structures are not all independent in $d=4$ due to Gram identities. The reason for considering six derivatives is that this is the most that can be generated from field redefinitions in a two-derivative theory. In any case, this limit on the number of derivatives in the contact term can be dropped, using the approach from \cite{Bonifacio:2018vzv,Bonifacio:2018aon}, and the final sum rules are the same.} With this, we can then find the full tree-level four-point scattering amplitude $h^\star h^\star \rightarrow h^\star h^\star$ and impose that it grows at most like $ E^2$ at high energies for all choices of external polarizations.  This gives a system of constraints on the cubic and quartic couplings.  The quartic couplings appear linearly and can be eliminated from this system. The resulting 43 constraints imply that certain sums of squares of linear combinations of the cubic couplings vanish, which using the reality of the couplings gives
\be
a_3  = 2 a_2, \quad b_1 =b_3, \quad 2 e_{2,a} = e_{3,a} = 2e_{4,a}= e_{5,a}, \quad a_1 = e_{1,a} = c_{2, \mathcal{I}} =d_{i} =0.
\ee
After imposing these conditions, we are left with the following three sum rules:  
\begin{subequations} \label{eq:sumrules}
\begin{align}
&a_2^2 +4 b_1^2 - \frac{3}{4} \sum_{a}  \left(\frac{m_{a}^2}{m_{\star}^2}-\frac{4}{3} \right) e_{5,a}^2=0\, , \label{eq:sum_rule_1} \\
&a_2^2+4 b_1^2 -24  \sum_{\mathcal{I}} c_{1, \mathcal{I}}^2+\frac{5}{4} \sum_{a}  \left(\frac{m_{a}^2}{m_{\star}^2}-\frac{4}{5} \right) \frac{m_a^2}{m_{\star}^2} e_{5,a}^2=0\, ,  \label{eq:sum_rule_2} \\
& 6 \sum_{\mathcal{I}} \frac{m_{ \mathcal{I}}^2}{m_{\star}^2} c_{1,\mathcal{I}}^2 + \frac{1}{4} \sum_{a} \left(\frac{m_{a}^2}{m_{\star}^2}-4\right)\left(\frac{m_{a}^2}{m_{\star}^2}-1\right) \frac{m_{a}^2}{m_{\star}^2} e_{5,a}^2 =0\, ,  \label{eq:sum_rule_3} 
\end{align}
\end{subequations}
which are given in Eq. (3.85) of \cite{Bonifacio:2019ioc}. 

\section{Scalar mass bounds from the sum rules}

We now proceed to derive the bound \eqref{mainbounde} from the sum rules. The argument follows the approach of the conformal bootstrap for bounding operator dimensions \cite{Rattazzi:2008pe, Rychkov:2009ij}, albeit with a small set of consistency conditions.
Consider a linear combination of the sum rules \eqref{eq:sumrules} defined by a constant vector $\vec{\alpha} = (\alpha_1, \alpha_2, \alpha_3)$, namely $\alpha_1$ times \eqref{eq:sum_rule_1}, plus $\alpha_2$ times \eqref{eq:sum_rule_2}, plus $\alpha_3$ times \eqref{eq:sum_rule_3}. We can find the $\vec{\alpha}$ that produce the strongest bounds using \texttt{SDPB} \cite{Simmons-Duffin:2015qma, Landry:2019qug}. 

First, we let $h_{\mu \nu}^{\star}$ be any massive spin-2 mode and consider the combination of sum rules given by $\vec{\alpha} =(1,1,0) $. This gives
\be
24  \sum_{\mathcal{I}} c_{1, \mathcal{I}}^2=2a_2^2 +8 b_1^2 + \frac{1}{4} \sum_{a} \left(5 \frac{m_{a}^4}{m_{\star}^4}  - 7 \frac{m_{a}^2}{m_{\star}^2}+ 4\right) e_{5,a}^2\,.
\ee
The assumption that the massive spin-2 mode $h_{\mu \nu}^{\star}$ has a non-vanishing four-point amplitude implies that at least one of the cubic couplings is non-vanishing,\footnote{A non-vanishing cubic coupling is guaranteed in the presence of gravity by the S-matrix equivalence principle, but in a non-gravitational theory it is an additional assumption.}  since pure contact interactions grow faster than $E^2$.
Since not all of the cubic couplings vanish, and because $5 x^2-7x+4>0$ for any $x$, both sides of this equation must be strictly positive. There must therefore be some non-vanishing scalar coupling $c_{1, \mathcal{I}}$ for every massive spin-2 mode. 

Now let $h_{\mu \nu}^{\star}$ be a massive spin-2 mode with the smallest mass, $m_{\star}=m_{1}$, and take $\vec{\alpha} =(-8,9,27) $. This gives
\be
a_2^2 +4  b_1^2+162 \sum_{\mathcal{I}} \left(\frac{m_{ \mathcal{I}}^2}{m_{\star}^2}-\frac{4}{3}\right) c_{1,\mathcal{I}}^2   +\frac{27}{4}  \sum_{a}   \left(\frac{4}{3} -  \frac{m_a^2}{m_{\star}^2}\right)^2 \left(  \frac{m_a^2}{m_{\star}^2}-\frac{2}{3}\right) e_{5,a}^2  =0\, .
\ee
For  $m_{\star}=m_{1}$ we have $m_a/m_{\star} \geq 1$, so all contributions are manifestly non-negative except for the sum over scalar modes. This implies that the scalar sum must be negative or zero. Since we already established that we cannot have all scalar couplings vanishing, there must be a scalar mode with nonzero coupling $c_{1,\mathcal{I}}$ to the lightest KK graviton and with $m_{ \mathcal{I}}^2/m_{1}^2 \leq 4/3$.  This proves the bound proposed in \cite{Mirbabayi:2026saz}. In the equality case, each term in the sum vanishes so gravity decouples, the self-coupling vanishes, and the lightest KK graviton couples only to scalars and heavier gravitons with masses $2 m_{1}/\sqrt{3}$, in agreement with \cite{Mirbabayi:2026saz}. If we have dynamical gravity, then $b_1 \neq 0$ and there must be a strictly negative contribution from the scalar sum, in which case the inequality is strict.

Lastly, we derive the new bound \eqref{newboundee}.   Let $h_{\mu \nu}^{\star}$ again be any massive spin-2 mode, with $m_{\star}=m_{n}$, and take $\vec{\alpha} =(0,9,25) $, which gives
\be
9a_2^2+36 b_1^2 +150 \sum_{\mathcal{I}} \left( \frac{m_{ \mathcal{I}}^2}{m_{\star}^2} -\frac{36}{25} \right)c_{1,\mathcal{I}}^2+\frac{25}{4} \sum_{a}  \left(\frac{8}{5} - \frac{m_{a}^2}{m_{\star}^2}\right)^2 \frac{m_{a}^2}{m_{\star}^2}  e_{5,a}^2  =0.
\ee
We have $m_a/m_{\star} > 0$, so all contributions are non-negative except for the sum over scalar modes. The contribution from the scalar sum must be negative or zero, so we learn that every massive spin-2 particle must have nonzero coupling to a scalar with $m_{ \mathcal{I}}^2/m_{n}^2 \leq 36/25$.  The inequality is actually strict, since saturation combined with  \eqref{eq:sum_rule_1} and  \eqref{eq:sum_rule_2} would imply that all cubic couplings vanish, which contradicts our assumption.

\section{Conclusions}

We have proven the bound \eqref{mainbounde} of Mirbabayi and Villadoro \cite{Mirbabayi:2026saz}, which gives an upper bound on the mass of a scalar particle that must exist and couple to the lowest massive spin-2 state in any theory with spins $\leq 2$ and parity-even two-derivative couplings, whose amplitudes grow no faster than $E^2$ with energy.  This includes KK theories, where the spin-2 state is the lightest massive KK graviton and the scalars include moduli fields. 
In this case, it bounds how strongly moduli can be stabilized: the moduli cannot all be stabilized with masses parametrically higher than the KK scale unless another light scalar couples to the lightest KK graviton. We have also proven the new bound \eqref{newboundee}, which applies to all massive spin-2 states.  Both bounds follow from sum rules on spin-2 scattering given in \cite{Bonifacio:2019ioc}, which can also be obtained from the constraints given in \cite{Mirbabayi:2026saz}.

A class of examples subject to these bounds is Freund--Rubin-type flux compactifications with a stable Minkowski vacuum \cite{Freund:1980xh}. Following \cite{Hinterbichler:2013kwa} (see also \cite{Brown:2013mwa}), if the internal space is an $N$-sphere, then stable 4D Minkowski solutions exist for $N=2$ and $N=3$. For $N=2$, the lightest scalar is the volume modulus and its mass satisfies  $m_{ \mathcal{I}}^2/m_{1}^2 = 1/2$. For $N=3$, the volume modulus gives a ratio of $16/15$, but the lightest scalar is a mixed mode coming from  the $l=2$ spherical harmonics and a scalar component of the higher-dimensional $2$-form with a mass-squared ratio of $m_{ \mathcal{I}}^2/m_{1}^2 =  8(6-\sqrt{31})/15 \approx 0.231$, which is comfortably below the bound \eqref{mainbounde}. 

It would be interesting to search for more bounds of this type that can be derived from the sum rules.
One immediate extension is to consider scattering in higher dimensions: here and in \cite{Mirbabayi:2026saz}, only theories in $d=4$ Minkowski space are considered. It would be interesting to see how the bound changes for $d\neq 4$.  
It would also be interesting if an analogous bound could be found for (anti) de Sitter space, using the consistency of dual CFT correlators.

\renewcommand{\em}{}
\bibliographystyle{utphys}
\addcontentsline{toc}{section}{References}
\bibliography{moduli_arxiv_v1}

\end{document}